\documentstyle[12pt,epsfig]{article} 
\begin{document}
\title{Space--time torsion contribution to quantum interference phases}
\author{ A. Camacho 
\thanks{email: acamacho@janaina.uam.mx}~ and 
A. Mac{\'\i}as \thanks{email: amac@xanum.uam.mx}\\
Physics Department, \\
Universidad Aut\'onoma Metropolitana-Iztapalapa. \\
P. O. Box 55-534, C. P. 09340, M\'exico, D.F., M\'exico.}
\date{}
\maketitle

\begin{abstract}
The field of neutron interferometry achieved one of its most significant su\-ccesses with the detection of the influence of gravity in the quantum mechanical phase of a thermal neutron beam. 
From the latest experimental readouts in this context an intriguing discrepancy has been elicited. Indeed, theory and ex\-periment dissent by one per cent, and though this fact could be a consequence of the mounting of the experimental device, it might also embody a difference between the way in which gravity behaves in classical and quantum mechanics. In this work the effects, upon the interference pattern, of space--time torsion will be analyzed heeding its coupling with the spin of the neutron beam. It will be proved that, even with this contribution, there is enough leeway for a further discussion of the validity of the equivalence principle in nonrelativistic quantum mechanics.
\end{abstract}
\bigskip
\bigskip

\section{Introduction}

The quantum mechanical phase, induced by gravity in a neutron interferometer, detected in 1975 by Colella, Overhauser, and Werner [1], spurred a series of experiments (usually known as COW), in which the involved interferometric techniques showed an increasing sophistication [2, 3]. This last fact opened up the possibility of testing the equivalence principle in the quantum realm resorting to a series of experiments, where the improvement of the accuracy thrived significantly [3]. 

All these experimental efforts finally paid off, since a disturbing discrepancy, on the order of one percent, between theory and experiment, emerged from the measurement readouts [4].
Clearly, a further analysis of the role of the equivalence principle in nonrelativistic quantum mechanics requires first the study of the consequences of some, not always taken into account, variables.  

For instance, in the case of a $1/2$--spin particle immersed in a Riemann--Cartan spacetime, how does the contribution to the interference pattern, stemming from the coupling spin--torsion, look like, as a function of the way in which neutron beam has been constructed? In other words, let us suppose that the spin part of the neutron beam's wave function is the coherent linear superposition of two contributions, one with z-component of the spin $1/2$, and the other one with $-1/2$. One question that could be posed at this point is the feasibility of the detection of the coupling spin--torsion looking at the changes that appear in the interference pattern as a function of the way in which the superposition is constructed.

At this point it is noteworthy to mention that, though, there are already some analyses of the consequences, in a interferometric experiment, of spacetime torsion, the aforesaid question has not been considered [5]. In the present work the effects, upon the interference pattern, of a contribution term stemming from torsion, are studied. However, here we prove that the presence of torsion could be detected, in principle, heeding the changes that appear as a function of the way in which the superposition is done. 

Additionally, it will be shown that the quantum mechanical trait of this effect depends on powers of $m/\hbar$, and hence has a striking similarity with its counterpart in the common COW experiment [1]. This dependence  has been understood, by some authors [6], as a possible quantum mechanical protrusion of a nongeometric feature of gravity, and therefore, bearing this remark in mind we may assert that nongeometri\-city pervades the movement of a quantum system immersed in a Riemann--Cartan manifold.
\bigskip
\bigskip

\section{Torsion and rotation in spin space}

Let us consider a neutron interferometer [1], and neglect the consequences of the earth's rotation on this kind of experimental construction [7, 8]. The main reason behind this approximation relies upon the fact that we would like to use this kind of approaches also as a test for the equivalence principle in the quantum realm. The analysis of rotation effects not only requires the locality hypothesis, in addition it imposes a generalization of this assumption, since it is necessary to know how accelerated observers measure wave characteristics in such a way that, in the eikonal limit, the recovery of the hypothesis of locality is ensured [8]. From the last argument it is readily seen that the introduction of rotation entails the presence of a cluster of assumptions that could cloud the final interpretation of our results.

The Hilbert space in this case is the tensor product of two contributions, to wit, spin state space, $\mathcal{E}$$_s$ and the orbital state space, $\mathcal{E}$$_r$. The dynamics of the state vector associated with the neutron beam will be described by the nonrelativistic limit of the Dirac equation, in a Newtonian approximation of Riemann--Cartan spacetime, namely, the Pauli equation [9]. 

{\setlength\arraycolsep{2pt}\begin{eqnarray}
i\hbar{\partial\vert\psi>\over\partial t} = -{\hbar^2\over 2m}\nabla^2\vert\psi> - i{\hbar^2\over m}\kappa_{(0)}\sigma^l\partial_l\vert\psi> - mV\vert\psi> - \hbar c\kappa_l\sigma^l\vert\psi>.
\end{eqnarray}}

In the foregoing expression the following terms have been considered, $c$ is the speed of light, $V$ the Newtonian gravitational potential, $\sigma^l$ Pauli matrices, and $\kappa_{\mu}$ the axial part of the spacetime torsion. Inasmusch as the rotation of the neutron interferometer has been neglected, we explain the absence of a coupling term between the interferometer and the Earth's rotation in this last equation. Additionally, in (1) we will consider that $\kappa_{(0)} = 0$. This simplification will allow us to fathom, in a clear manner, the consequences, upon the interefometric pattern, of the space part of the axial part of the torsion. 

{\setlength\arraycolsep{2pt}\begin{eqnarray}
i\hbar{\partial\vert\psi>\over\partial t} = -{\hbar^2\over 2m}\nabla^2\vert\psi> - mV\vert\psi> - \hbar c\kappa_l\sigma^l\vert\psi>.
\end{eqnarray}}

  Hence, denoting by $\phi$ the spin state vector, we find that its dynamics is governed by

{\setlength\arraycolsep{2pt}\begin{eqnarray}
i\hbar{\partial\phi\over\partial t} = -\hbar c\kappa_n\sigma^n\phi.
\end{eqnarray}}

It is readily seen that the solution reads

{\setlength\arraycolsep{2pt}\begin{eqnarray}
\phi(t) = \exp\Bigl\{ic\int_0^t\kappa_n\sigma^ndt'\Bigr\}\phi(t= 0).
\end{eqnarray}}

Let us now consider the case in which we perform an
experiment similar to COW [1], i.e., two particles, starting
at point $(O)$, move along two different trajectories, $C$ and $\tilde{C}$, and afterwards
they are detected at a certain point $S$. Here we assume that the size of the
wavelengths of the packets is much smaller than the size in which the field changes considerably (i.e., we are always in the short wavelength limit), and in consequence we may consider a semiclassical approach in the analysis of the wave function. Trajectory $C$ is made up of two contributions, namely, (O)--(A) which is horizontal, whose length reads $l$, and (A)--(S), vertical, and with length equal to $L$. $\tilde{C}$ comprises also two parts, (O)--(B) vertical, with length $L$, and (B)--(S) horizontal, and size $l$. The horizontal axis is $x$, and $y$ points upwards, such that the Newtonian potential reads $V = gy$.

 Additionally, we assume that 

{\setlength\arraycolsep{2pt}\begin{eqnarray}
\kappa_n(A) = \kappa_n(0) + {\partial\kappa_n\over\partial x}_{(0)}l,
\end{eqnarray}}

{\setlength\arraycolsep{2pt}\begin{eqnarray}
\kappa_n(B) = \kappa_n(0) + {\partial\kappa_n\over\partial y}_{(0)}L.
\end{eqnarray}}

Hence, it is deduced that at the screen, (S), (for the spin wave function that passes through (A), $\phi_A(S)$, and for that passing through (B), $\phi_B(S)$) we have

{\setlength\arraycolsep{2pt}\begin{eqnarray}
\phi_A(S) = \exp\Bigl\{ic\sigma^n[\alpha_A\kappa_n(0) + \beta_A{\partial\kappa_n\over\partial x}_{(0)} + \gamma_A{\partial\kappa_n\over\partial y}_{(A)}]\Bigr\}\phi(t= 0),
\end{eqnarray}}

{\setlength\arraycolsep{2pt}\begin{eqnarray}
\phi_B(S) = \exp\Bigl\{ic\sigma^n[\alpha_B\kappa_n(0) + \beta_B{\partial\kappa_n\over\partial x}_{(B)} + \gamma_B{\partial\kappa_n\over\partial y}_{(0)} ]\Bigr\}\phi(t= 0).
\end{eqnarray}}

In these last two expressions we have (approximately)

{\setlength\arraycolsep{2pt}\begin{eqnarray}
\alpha_A = {m\tilde{\lambda}\over\hbar}\Bigl\{l + L/2 - ({m\tilde{\lambda}\over\hbar})^2gL^2/8\Bigr\},
\end{eqnarray}}

{\setlength\arraycolsep{2pt}\begin{eqnarray}
\beta_A = {m\tilde{\lambda}\over\hbar}l\Bigl\{(l + L)/2 - ({m\tilde{\lambda}\over\hbar})^2gL^2/8\Bigr\},
\end{eqnarray}}

{\setlength\arraycolsep{2pt}\begin{eqnarray}
\gamma_A = {m\tilde{\lambda}\over\hbar}\Bigl\{L^2/2\Bigl[1/4 - ({m\tilde{\lambda}\over\hbar})^2gL/4\Bigr] + lL\Bigl[1/2 - ({m\tilde{\lambda}\over\hbar})^2g(2L + 3l)/4\Bigr]\Bigr\},
\end{eqnarray}}

{\setlength\arraycolsep{2pt}\begin{eqnarray}
\alpha_B = {m\tilde{\lambda}\over\hbar}\Bigl\{l + L/2 + ({m\tilde{\lambda}\over\hbar})^2gL\Bigl[l - L/8\Bigr] \Bigr\},
\end{eqnarray}}

{\setlength\arraycolsep{2pt}\begin{eqnarray}
\beta_B = 3L^2{m\tilde{\lambda}\over\hbar}\Bigl\{1/4 - ({m\tilde{\lambda}\over\hbar})^2gL/8\Bigr\},
\end{eqnarray}}

{\setlength\arraycolsep{2pt}\begin{eqnarray}
\gamma_B = {m\tilde{\lambda}\over\hbar}L^2\Bigl\{3/4\ - ({m\tilde{\lambda}\over\hbar})^213gL/(48)\Bigr\}.
\end{eqnarray}}

In all our equations $\tilde{\lambda} = \lambda/(2\pi)$, and $\lambda$ denotes the initial wavelength of the neutron beam.

These two wave functions may be written in terms of a rotation of the initial state

{\setlength\arraycolsep{2pt}\begin{eqnarray}
\phi_n(S) = \exp\Bigl\{-{i\over 2}\theta_v\vec{n}_v\cdot\vec{\sigma}\Bigr\}\phi(t= 0).
\end{eqnarray}}

Here $v = A, B$. The definition of the components of the unit vectors and the rotation angles are given by

{\setlength\arraycolsep{2pt}\begin{eqnarray}
\tau_n^{(A)} = \Bigl\{\alpha_A \kappa_n(0) + \beta_A{\partial\kappa_n\over\partial x}_{(0)}l+ \gamma_A{\partial\kappa_n\over\partial y}_{(A)}\Bigr\}, 
\end{eqnarray}}

{\setlength\arraycolsep{2pt}\begin{eqnarray}
(\vec{n}_{A})_n = {\tau_n^{(A)}\over\sqrt{(\tau_x^{(A)})^2 + (\tau_y^{(A)})^2 + (\tau_z^{(A)})^2}},
\end{eqnarray}}

{\setlength\arraycolsep{2pt}\begin{eqnarray}
\theta_A = -2c\sqrt{(\tau_x^{(A)})^2 + (\tau_y^{(A)})^2 + (\tau_z^{(A)})^2}.
\end{eqnarray}}

Likewise for case (B).

>From our results we may distinguish two different situations: 
\bigskip

(i) $\vert l{\partial\kappa_n\over\partial y}\vert, \vert l{\partial\kappa_n\over\partial x}\vert << \vert\kappa_n\vert$. Therefore $\vec{n}_A = \vec{n}_B$, the axis of rotation of the beams is the same, and they differ only in the angle of rotation, $\theta_A \not= \theta_B$.
\bigskip

(ii) Whereas if the foregoing condition does not hold, then not only $\theta_A \not= \theta_B$, but additionally $\vec{n}_A \not= \vec{n}_B$.
\bigskip
\bigskip

\section{Interference patterns and superposition of quantum states}
\bigskip

\subsection{General case}

Let us now assume that $\phi(t= 0)$ is the linear coherent superposition of states $\chi_{(+)}$ and $\chi_{(-)}$, where $\sigma_z\chi_{(\pm)} = \pm\chi_{(\pm)}$, namely

{\setlength\arraycolsep{2pt}\begin{eqnarray}
\phi(t= 0) = c_{(+)}\chi_{(+)} + c_{(-)}\chi_{(-)}.
\end{eqnarray}}

The interference pattern at $S$ is a function of the complete state vector, i.e., $\vert\psi>$, whose dynamics evolves according to (1). We may rephrase this last argument stating $I = \vert(\vert\psi>_{(A)} + \vert\psi>_{(B)})\vert^2$, and it comprises two different contributions, one stemming from $\mathcal{E}$$_s$ and the second one from $\mathcal{E}$$_r$. In other words, we find that

{\setlength\arraycolsep{2pt}\begin{eqnarray}
I = 2 + 2\cos\Bigl(({m\over\hbar})^2glL\tilde{\lambda}\Bigr)\Bigl[\phi\dagger_A(S)\phi_B(S) + \phi\dagger_B(S)\phi_A(S)\Bigr].
\end{eqnarray}}

Taking into account our previous definitions we have that

{\setlength\arraycolsep{2pt}\begin{eqnarray}
I = 2 + 2\cos\Bigl(({m\over\hbar})^2glL\tilde{\lambda}\Bigr)\Bigl[\cos({\theta_A\over 2})\cos({\theta_B\over 2}) + [\vec{n}_A\cdot\vec{n}_B]\sin({\theta_A\over 2})\sin({\theta_B\over 2})\Bigr] \nonumber\\
- 2\sin\Bigl(({m\over\hbar})^2glL\tilde{\lambda}\Bigr)\Bigl[\sin({\theta_A\over 2})\sin({\theta_B\over 2})[\vec{n}_A\times\vec{n}_B] + \sin({\theta_A\over 2})\cos({\theta_B\over 2})\vec{n}_A \nonumber\\
-\sin({\theta_B\over 2})\cos({\theta_A\over 2})\vec{n}_B\Bigr]\cdot\Bigl[2Re(c^{\ast}_{(+)}c_{(-)})\vec{e}_x - 2Im(c^{\ast}_{(-)}c_{(+)})\vec{e}_y  \nonumber\\
+ (\vert c_{(+)}\vert^2 - \vert c_{(-)}\vert^2\vec{e}_z\Bigr].
\end{eqnarray}}

Clearly, $\cos\Bigl(({m\over\hbar})^2glL\tilde{\lambda}\Bigr)$ corresponds to the interference term in COW [1, 6]. This means that if we discard torsion, then we recover COW. Additionally, $\vec{e}_n$ denotes the unit vector along the $n$--axis.
\bigskip
\subsection{Particular cases}

\bigskip
\subsubsection{$c_{(+)} = c_{(-)} = 1/\sqrt 2$}

Under these conditions we have that

{\setlength\arraycolsep{2pt}\begin{eqnarray}
I = 2 + 2\cos\Bigl(({m\over\hbar})^2glL\tilde{\lambda}\Bigr)\Bigl[\cos({\theta_A\over 2})\cos({\theta_B\over 2}) + [\vec{n}_A\cdot\vec{n}_B]\sin({\theta_A\over 2})\sin({\theta_B\over 2})\Bigr]. 
\end{eqnarray}}

\bigskip
\subsubsection{$c_{(+)}, c_{(-)} \in\Re$}

Here we consider $c_{(+)} \not = c_{(-)}$.
\bigskip

{\setlength\arraycolsep{2pt}\begin{eqnarray}
I = 2 + 2\cos\Bigl(({m\over\hbar})^2glL\tilde{\lambda}\Bigr)\Bigl[\cos({\theta_A\over 2})\cos({\theta_B\over 2}) + [\vec{n}_A\cdot\vec{n}_B]\sin({\theta_A\over 2})\sin({\theta_B\over 2})\Bigr] \nonumber\\
- 2\sin\Bigl(({m\over\hbar})^2glL\tilde{\lambda}\Bigr)\Bigl[\sin({\theta_A\over 2})\sin({\theta_B\over 2})[\vec{n}_A\times\vec{n}_B] + \sin({\theta_A\over 2})\cos({\theta_B\over 2})\vec{n}_A \nonumber\\
-\sin({\theta_B\over 2})\cos({\theta_A\over 2})\vec{n}_B\Bigr]\cdot[\vert c_{(+)}\vert^2 - \vert c_{(-)}]\vert^2\vec{e}_z.
\end{eqnarray}}

If, additionally, we neglect all derivatives of the axial part of the torsion, a condition that implies $\vec{n}_A = \vec{n}_B$, we obtain

{\setlength\arraycolsep{2pt}\begin{eqnarray}
I = 2 + 2\cos\Bigl(({m\over\hbar})^2glL\tilde{\lambda}\Bigr)\cos\Bigl(({m\tilde{\lambda}\over\hbar})^3gcl^2K\Bigr)  \nonumber\\
- 2\kappa(0)_z/K\Bigl[\vert c_{(+)}\vert^2 - \vert c_{(-)}\vert^2\Bigr]\sin\Bigl(({m\over\hbar})^2glL\tilde{\lambda}\Bigr)\sin\Bigl(({m\tilde{\lambda}\over\hbar})^3gcl^2K\Bigr).
\end{eqnarray}}

In the foregoing expression the following definition has been introduced $K = \sqrt{\kappa^2(0)_x+ \kappa^2(0)_y+ \kappa^2(0)_z}$.
\bigskip

\section{Conclusions}

Expression (21) allows us enough leeway to consider the possibility of detecting the consequences of torsion, upon the interference pattern, modifying the values of $c_{(+)}$ and $c_{(-)}$. For instance, choosing $c_{(+)} = 1$ and $c_{(-)} = 0$, 
\bigskip

{\setlength\arraycolsep{2pt}\begin{eqnarray}
I = 2 + 2\cos\Bigl(({m\over\hbar})^2glL\tilde{\lambda}\Bigr)\cos\Bigl(({m\tilde{\lambda}\over\hbar})^3gcl^2K\Bigr)  \nonumber\\
- 2\kappa(0)_z/K\sin\Bigl(({m\over\hbar})^2glL\tilde{\lambda}\Bigr)\sin\Bigl(({m\tilde{\lambda}\over\hbar})^3gcl^2K\Bigr).
\end{eqnarray}}

Resorting now to $c_{(+)} = 0$ and $c_{(-)} = 1$

{\setlength\arraycolsep{2pt}\begin{eqnarray}
I = 2 + 2\cos\Bigl(({m\over\hbar})^2glL\tilde{\lambda}\Bigr)\cos\Bigl(({m\tilde{\lambda}\over\hbar})^3gcl^2K\Bigr)  \nonumber\\
+ 2\kappa(0)_z/K\sin\Bigl(({m\over\hbar})^2glL\tilde{\lambda}\Bigr)\sin\Bigl(({m\tilde{\lambda}\over\hbar})^3gcl^2K\Bigr).
\end{eqnarray}}

As we switch from $\{c_{(+)} = 1, c_{(-)} = 0\}$ to $\{c_{(+)} = 0, c_{(-)} = 1\}$ a sign change, in the second term of the right--hand side, emerges. This effect disappears if torsion vanishes. In other words, this sign change is a direct consequence of torsion, and appears only if we modify the linear superposition of the starting spin state vector. As a matter of fact, considering a series of experiments, in which we begin with $\{c_{(+)} = 1, c_{(-)} = 0\}$, and gradually we change these two values (the first parameter diminishes, whereas the second one increases), then the role, that the absolute value of the second term plays, would peter out, this happens when $c_{(+)} = 1/\sqrt2$. Afterwards, it starts to appear, once again.

Let us now estimate the order of magnitude of the torsion contributions, and afterwards confront them with the current experimental discrepancy. To circunvent all possible encumbrance in the physical analysis we will assume that $c_{(+)} = 1$ and  $\kappa(0)_z/K = 1$. In this way

{\setlength\arraycolsep{2pt}\begin{eqnarray}
I = 2\Bigl\{1 + \cos\Bigl(({m\over\hbar})^2lg\tilde{\lambda}[L + {m\over\hbar}cl\tilde{\lambda}^2K]\Bigr)  \Bigr\}.
\end{eqnarray}}

The theoretical result, no torsion included [6], shows a discrepancy on the order of one percent in the phase shift [3]. Denoting the contribution to this discrepancy, stemming from torsion, with $\Gamma$, we have

{\setlength\arraycolsep{2pt}\begin{eqnarray}
({m\over\hbar})^2lg\tilde{\lambda}[L + {m\over\hbar}cl\tilde{\lambda}^2K]\Bigr) = ({m\over\hbar})^2lg\tilde{\lambda} L[1 + \Gamma].
\end{eqnarray}}

The most stringent experimental bound reads $K\sim 10^{-15}m^{-2}$ [9], and hence (employing the typical experimental values [1, 3, 6]), we deduce 

{\setlength\arraycolsep{2pt}\begin{eqnarray}
\Gamma\sim 10^{-16}.
\end{eqnarray}}

Firstly, one of the conclusions to be drawn from (29) comprises the assertion that the involved experimental discrepancy can not be fathomed resorting, exclusively, to torsion effects, and in consequence, there is enough leeway to continue the discussion around the validity of the equivalence principle in the quantum realm [10].

Secondly, as already known [6], the appearance of the mass term in the interference expression ($[{m\over\hbar}]^2$) has been understood by some authors as a possible manifestation of nongeometricity in the gravitational field. Taking a look at (24) it is readily seen that under the aegis of torsion this trait, not only does not vanish, but on it an additional term is bestowed, i.e., $[{m\over\hbar}]^3$.  Therefore, bearing this remark in mind we may assert that nongeometricity pervades the movement of a nonrelativistic quantum system immersed in a Riemann--Cartan manifold.

\bigskip
\bigskip

\Large{\bf Acknowledgments.}\normalsize
\bigskip

This contribution is dedicated to Jerzy Pleba\'nski, whose toil in Physics evinces his wit and perseverance, virtues that have guided several generations of physicists. The present work was supported by CONACyT Grant 42191--F.

\end{document}